\documentclass[article,14pt,onecolumn,showpacs]{revtex4-1}

\usepackage{graphicx}
\usepackage{amsfonts}
\usepackage{amsmath}
\usepackage{amssymb}
\usepackage{color,ulem}
\usepackage{chngcntr}

\newcommand{\removed}[1]{}

\begin{document}

\title{Microwave-stimulated superconductivity due to presence of vortices}
\author{Antonio Lara$^{1}$}
\author{Farkhad G. Aliev$^{1}$}
\email{Corresponding author: farkhad.aliev@uam.es}
\author{Alejandro V. Silhanek$^{2}$}
\author{Victor V. Moshchalkov$^{3}$}

\bigskip

\affiliation{$^{1}$Dpto. Fisica Materia Condensada, Instituto Nicolás Cabrera (INC) and Condensed Matter Physics Institute (IFIMAC), Universidad Aut{\'o}noma
de Madrid, 28049, Madrid, Spain}

\affiliation{$^{2}$INPAC- Katholieke Universiteit Leuven, Celestijnenlaan 200D,
B3001, Leuven, Belgium and Departement de Physique, Universit{\'e} de Li{\`e}ge, B-4000 Sart Tilman, Belgium}

\affiliation{$^{3}$INPAC- Katholieke Universiteit Leuven, Celestijnenlaan 200D,
B3001, Leuven, Belgium}

\date{\today }

\maketitle

\textbf{The response of superconducting devices to electromagnetic radiation is a core concept implemented in diverse applications, ranging from the currently used voltage standard to single photon detectors in astronomy. Suprisingly, a sufficiently high power subgap radiation may stimulate superconductivity itself. The possibility of stimulating type II superconductors, in which the  radiation may interact also with vortex cores, remains however unclear. Here we report on superconductivity enhanced by GHz radiation in type II superconducting Pb films in the presence of vortices. The stimulation effect  is more clearly observed in the upper critical field and less pronounced in the critical temperature. The magnetic field dependence of the vortex related microwave losses in a film with periodic pinning reveals a reduced dissipation of mobile vortices in the stimulated regime due to a reduction of the core size. Results of numerical simulations support the validy of this conclusion.
  Our findings may have intriguing connections with holographic superconductors in which the possibility of stimulation is under current debate.}

Microwave (\textit{mw}) irradiation has been used to control the quantum properties of different systems, from supercurrents in superconductors to mechanical oscillators \cite{Lindner2011,McIver2011,Bergeret2010,Palomaki2013}. Using nonequilibrium pumping for cooling is currently a hot topic \cite{cooling_barrier, cooling_phase}. In 1966 microwave stimulated superconductivity (MSSC) was discovered \cite{Wyatt1966} in superconducting bridges and later confirmed for different type I superconducting systems such as films \cite{Pals1979,Tolpygo1983}, tunnel junctions \cite{Heslinga1993} or cylinders \cite{Pals_Cylinder,Response_to_Pals}. This counterintuitive effect was explained by Eliashberg~\cite{Eliashberg1970} as a consequence of an irradiation-induced redistribution of quasiparticles away from the gap edge. Very recently MSSC has been observed in transient regimes (on $ps$ time scales) in NbN films \cite{Beck2013} and was demonstrated to improve the quality factor of superconducting \textit{mw} resonators \cite{deVisser2014}.

In type II superconductors with magnetic field penetrating in form of quantized flux (vortices) \cite{Abrikosov_paper}, the reduced inelastic relaxation time could suppress or modify some signatures of MSSC. A renewed interest in the type II SCs is related with the proposal of holographic superconductors (HS) \cite{Horowitz_PRL}, mapping solutions of astrophysics problems to scalar condensates. Just like as type II SCs in solid state, HSs can exhibit vortex configurations \cite{Barcelona} and can be phenomenologically described by the time dependent Ginzburg-Landau equation (TDGL) \cite{TDGL_HS} in the proximity of the critical temperature. The possibility of stimulated SC in the HSs is under current debate \cite{Enhanced_HS_1,Enhanced_HS_2}. Clearly, the experimental verification of stimulated superconductivity in type II SCs in the vortex state could therefore have important implications both inside and outside condensed matter physics community, paving the way for further progress in modelling 
 physics of black holes and gravity through HSs.

A periodic \textit{mw} pump of sufficient amplitude induces the motion of vortices that results in dissipation \cite{Gittleman1966,Silva1991,Golosovsky1996,Wordenweber2012}. Though dynamics of vortices was extensively addressed \cite{Blatter}, the possibility of MSSC in the vortex state is not fully understood. One can speculate that the energy balance in microwave-driven vortices should depend on a competition between friction-induced heating of quasiparticles in the vortex cores \cite{Clem1968,Shekhter2011,Gurevich2008} and energy pumping outside the core at large vortex velocities~\cite{Larkin1975}. However, the full picture of nonlinear electromagnetic response of the vortex matter in the proximity to the critical temperature remains unsettled. 

Our paper reports on broadband nonlinear response to \textit{mw} radiation in the GHz range in type II superconducting Pb films. We observe experimentally MSSC in an enhancement of the critical temperature, while much larger effects are seen in the second critical field. In order to investigate stimulation under varying pinning strength (i.e. vortex motion amplitude), we have carried detailed studies of MSSC in the Pb films with periodic pinning centers (Pb-PCC) and with applied DC field inclined about 4$^{\circ}$ off the film plane. Such configuration has been chosen because the most pronounced effect has been observed with it. Besides, the small perpendicular field component $H_{\perp }$ creates vortices, permitting us, by properly chosing the magnetic field intensity, to achieve the situations where the number of vortices is an integer multiple of the number of pinning centers (matching conditions). In that case, the vortex lattice rearranges itself in a specially stable manner, and vortex motion becomes restricted. Therefore, the commensurability between the vortex lattice and the periodic pinning centers facilitates the investigation of the complex vortex dynamics under periodically varying pinning conditions. The experiments unambiguously reveal a reduction of the dissipation of microwave driven mobile superconducting vortices at moderate frequencies and powers. Supported by TDGL simulations, we relate this unexpected behaviour to the reduction of the vortex core size at large vortex velocities, predicted by Larkin and Ovchinnikov (LO) ~\cite{Larkin1975} and seen indirectly when the flux is driven with a DC current \cite{Doettinger1994}.

\section*{Results}

\textbf{Stimulation of critical temperature and upper critical field.} A description of the samples and measurement details are provided in the Methods section and in the Supplementary material. Figure 1b shows the real and imaginary parts of the microwave
permeability parameter $U$, defined in the Methods section, measured in Pb-PPC at small \textit{mw} powers. Measurements are done in a temperature ($T$) sweep for different magnetic fields, at a fixed frequency ($f$) and are in accordance with the Coffey-Clem model \cite{Coffey1992}. The dependence of the response on the \textit{mw} power ($P$) for the same Pb-PPC sample is seen in the contour plot for $U^{\prime }$ in the plane $P-T$ for magnetic field $H=0$ and $f=6$ GHz, (Fig. 1c). To characterize the shift of the transition as a function of $P$, we introduce an effective critical
temperature, $T_{c}^{\ast }$, (Supplementary Fig. A3), determined with an error around 1 mK. Figure 1d presents a 3D plot of $%
T_{c}^{\ast }$ in the coordinates $P-H$ in the Pb-PPC at $f=6 $~GHz. One observes a non-monotonic dependence of $%
T_{c}^{\ast }(P)$, increasing at small $P$ and decreasing at large $P$.
Though the $T_{c}^{\ast }$ does not have the meaning of critical
temperature of the superconducting transition, the observed increase of $%
T_{c}^{\ast }$ can be interpreted as an indication of MSSC.

\begin{figure}[h]
\includegraphics[scale=0.3]{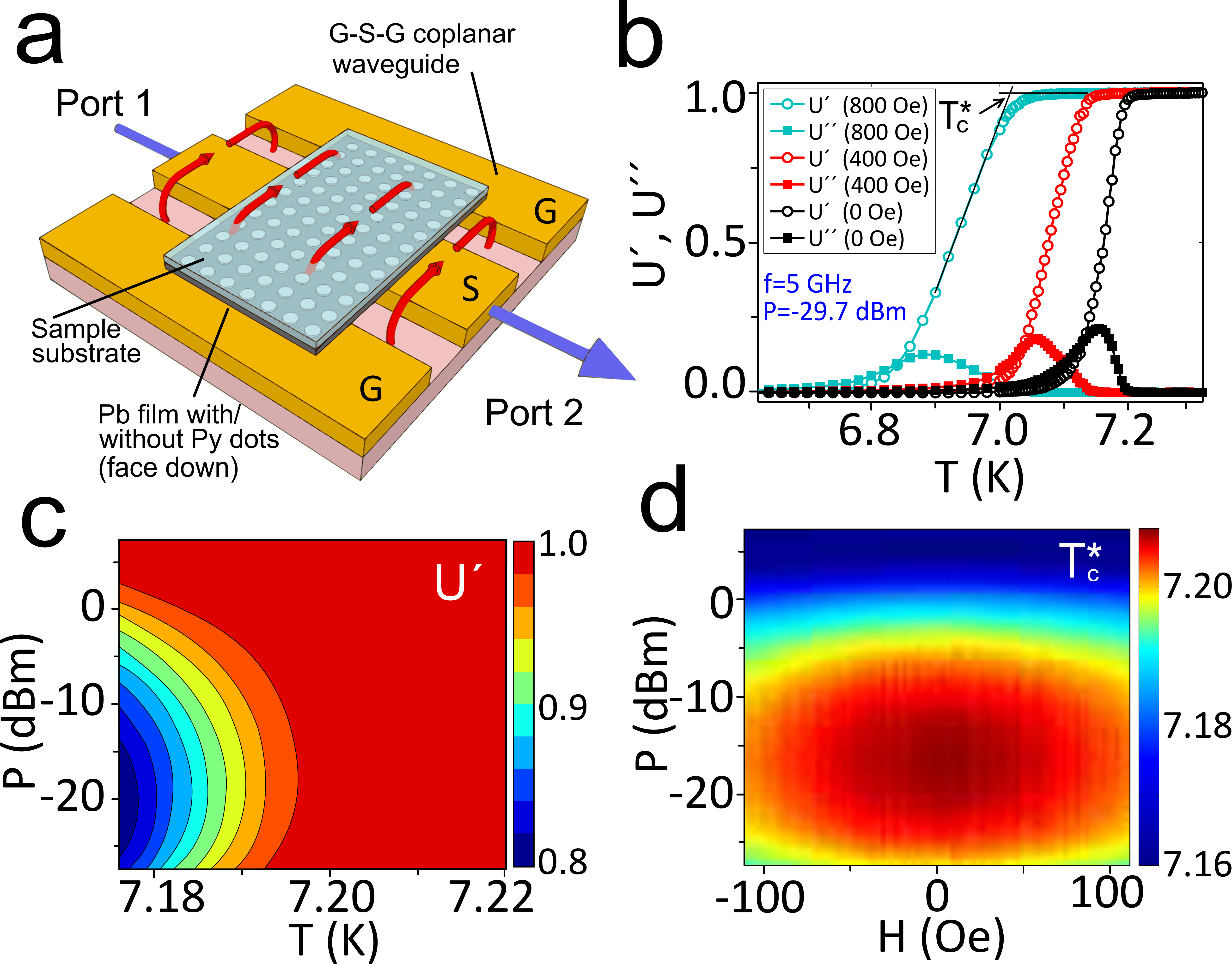}
\caption{Sketch of the sample and effect of stimulation on critical temperature.
a) Sketch of the sample placed face down on the
ground-signal-ground (G-S-G) coplanar waveguide (CPW). Red arrows represent the magnetic field generated by the CPW. b)
Measurements of $U^{\prime }$ and $U^{\prime \prime }$ at different fields.
c) Contour plot of $U^{\prime }$ versus microwave power and temperature,
measured at $f=6$ GHz. d) $T_c^*$ in the field range from $H=-120$ Oe to $%
H=120$ Oe. All data correspond to the Pb-PPC sample.}
\end{figure}

To investigate the nonlinear response in the vortex state as a function of pinning strength, we have carried out measurements of $U^{\prime }$ and $U^{\prime \prime }$ as
functions of $H$, $P$, $f$ and $T$ in Pb-PPC. Figure 2a shows a 3D plot of $U^{\prime }$ for the Pb-PCC  sample 
at $T=7.19$~K in the coordinates $H-P$. Dark red tones correspond to the normal state. As $H$ is lowered, the samples become superconducting and
the magnetic permeability changes in agreement with expectations 
\cite{Coffey1992}. Figure 2b shows a typical set of cross-sections of 
$U^{\prime }$ at fixed magnetic fields. While \textit{mw}
power above 5 dBm destroys superconductivity, for intermediate values the
superconducting response is the most intense. We will refer to the applied $P$ that yields the strongest superconducting response as \textquotedblleft optimum power" ($P_{%
\mathrm{O}}$). The dashed line indicates how $P_{\mathrm{O}}$ changes with $%
H$. The response (inset of Fig. 2b) shows directly transition between linear to nonlinear vortex response regimes. At lowest powers it flattens and is noisier , since the relative noise is larger compared to the weaker signal received in port 2 of the network analyzer.

Similarly to $T_{c}^{\ast }$, we introduce an effective critical magnetic field, $H_{c2}^{\ast}$. Figure 2c shows a plot of $H_{c2}^{\ast }(P,T)$ in the Pb-PPC sample (see Suplementary material for the method of finding $H_{c2}^{\ast}$, and the results for the plain film). For each temperature, $H_{c2}^{\ast}$ has been normalized by its value at the lowest $P$, so relative values of $H_{c2}^\ast$ can be compared. As $%
T\rightarrow T_{c}$, the relative increase of $H_{c2}^{\ast }$ under $mw$ at 
$P=P_{\mathrm{O}}$ becomes larger. Figure 2d compares the normalized $%
H_{c2}^{\ast}$ as a function of reduced temperature in both types of samples (with and without pinning centers),
showing that MSSC effects in $H_{c2}^{\ast }$ are stronger in the Pb-PPC
sample.

\begin{figure}
\includegraphics[scale=0.3]{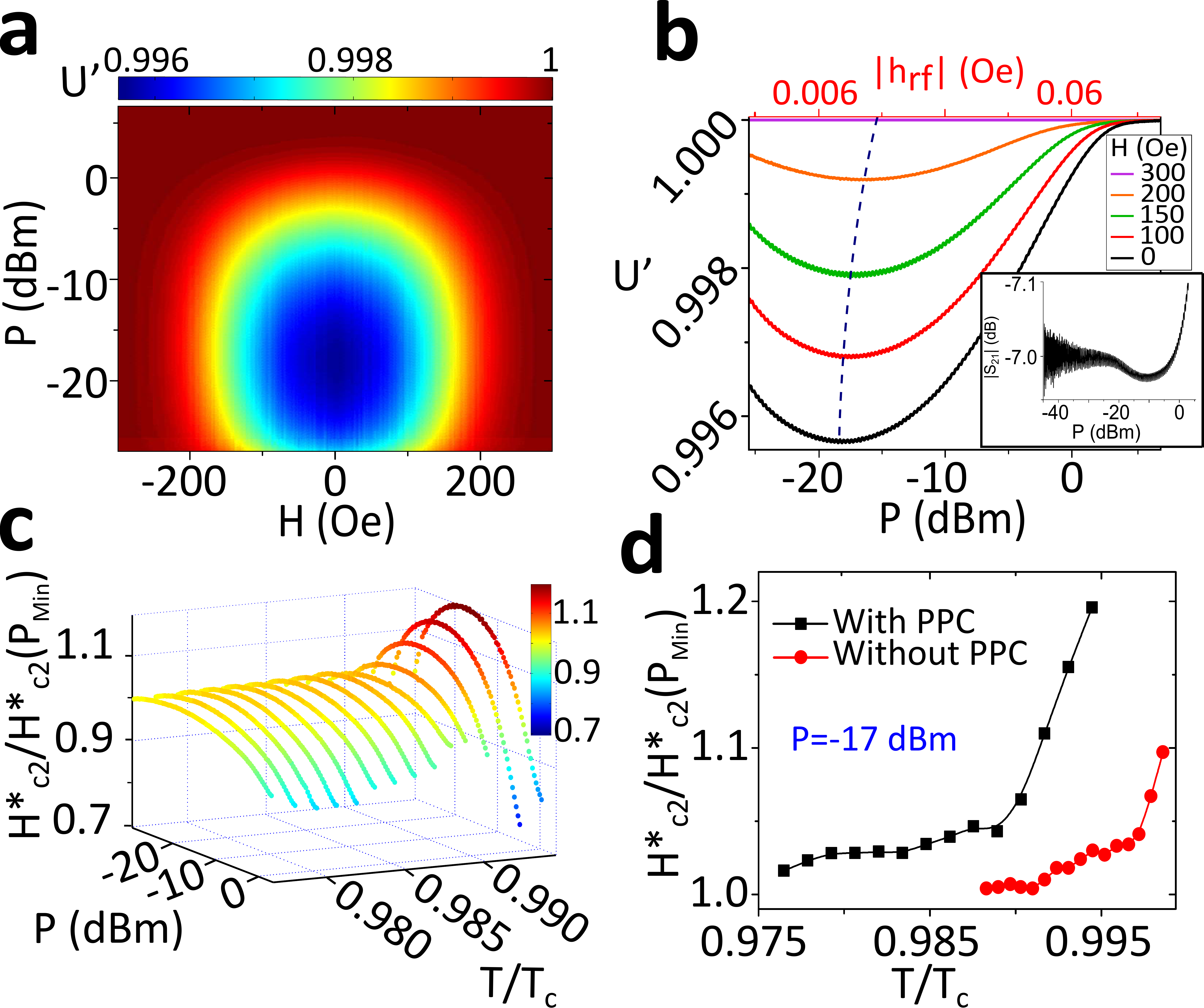}
\caption{Effect of stimulation on the upper critical field.
a) Real part of the microwave permeability at $T=7.19K$, as a
function of applied field and microwave power. b) Cross sections of panel a)
at fixed fields. The dashed line indicates optimum power. The inset shows a $\vert S_{21} \vert$ trace in a broader power range, at $H=0$ Oe, $f=4$ GHz and $T/T_c=0.991$. c) $H_{c2}^*$ for
different $T$ and $P$ for the Pb-PPC sample. Values are normalized at each $%
T $ by the values of $H_{c2}^*$ at the minimum power. d) Comparison of normalized $%
H_{c2}^* $ for each sample.}
\end{figure}

Figure 3a shows that $P=P_{\mathrm{O}}$ increases with \textit{mw} frequency with a maximum value that saturates around 15 GHz. These frequencies are well below those corresponding to the superconducting gap.

\textbf{Reduced dissipation of microwave driven mobile vortex.} The changes in $P_{\mathrm{O}}$ indicate that the ``\textit{cooling}" effectiveness of \textit{mw} radiation depends on the pinning strength through the applied field which changes the number of vortices per pinning center and correspondingly their mobility. In Fig. 3b this fact is exposed for $f=6$~GHz, at different values of $T$.  Solid lines show $P_{\mathrm{O}}$ vs. magnetic field with matching conditions indicated by vertical dotted lines. Red circles represent a typical measurement of $U^{\prime }(f,P,H)$ at fixed values of $f $ and $P$, in which  the same matching anomalies appear in the \textit{mw} permeability. The lowest value of  $P_{\mathrm{O}}$ for every temperature is found always at zero field, and decreases locally in matching conditions. This hints the relevance of the vortex mobility for the value of $P_{\mathrm{O}}$: the enhanced vortex mobility out of matching conditions provides relatively larger (respect to matching) $P_{\mathrm{O}}$ values and correspondingly larger "cooling" efficiency.

This counterintuitive result has been corroborated through a set of independent experiments investigating the $H$ and $P$ dependencies of $U^{\prime
\prime }$.  At low $f$, when MSSC is not yet pronounced, the dissipation ($U^{\prime
\prime }$) at matching conditions shows (as expected) dips in a
broad range of $P$ (Supplementary Fig. A.7). However, for higher $f$ the dips of losses at matching
conditions convert into peaks. The same effect can be observed as a function of power, in Fig. 3c. In other words,\textit{ vortices moving with higher average velocities out of matching conditions manage to dissipate less than pinned in matching conditions}. These observations indicate a qualitative
change in the microwave response of superconducting vortices at high $mw$ frequencies.

\begin{figure}
\includegraphics[scale=0.3]{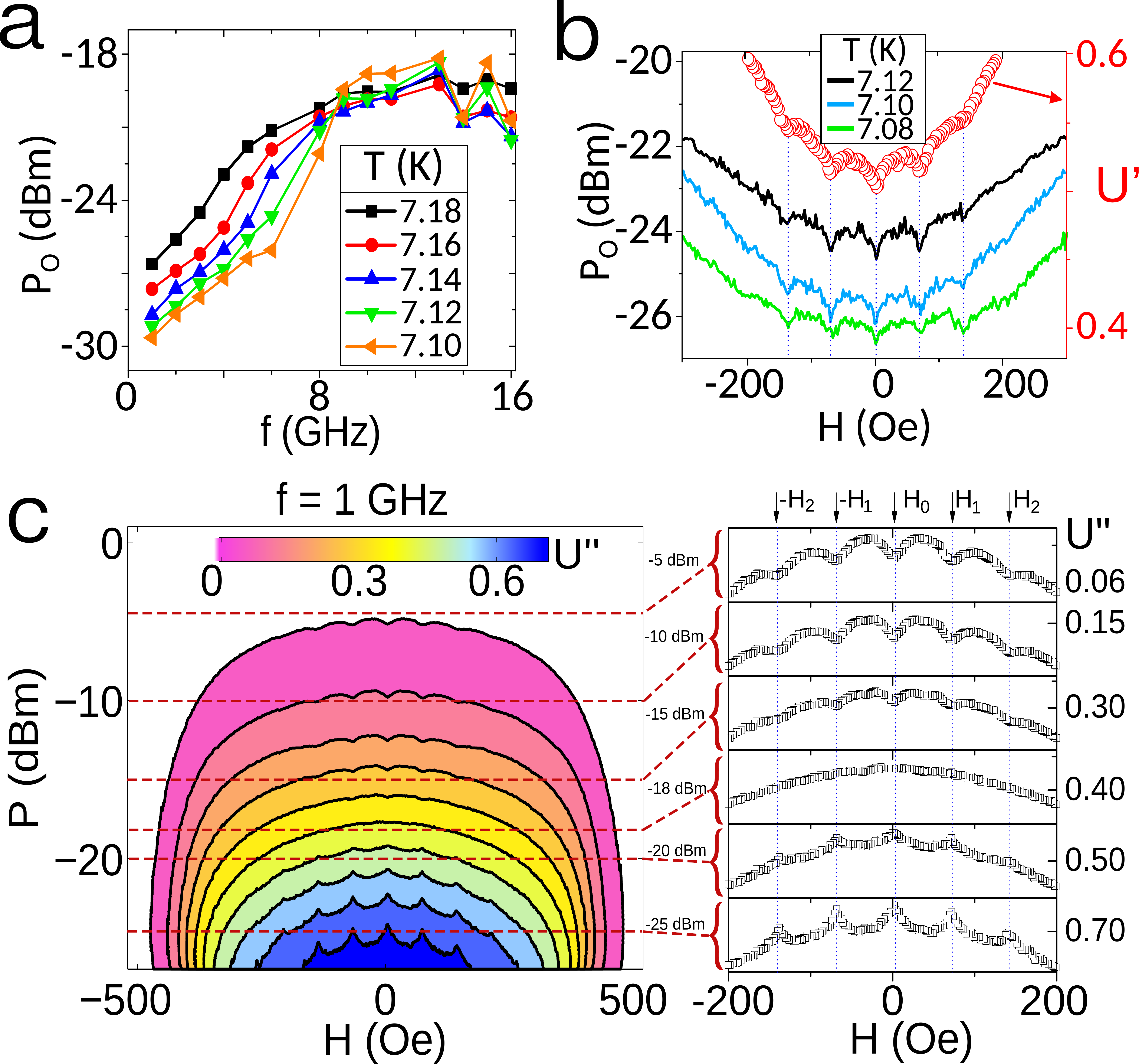}
\caption{Nonlinear response of vortex losses at matching conditions.
a) $P_{\mathrm{O}}$ versus $f$. b) Lines represent optimum power ($%
P_{\mathrm{O}}$) as a function of applied field at $%
f=6$~GHz. Red circles (right axis) represent $U^{\prime }$. Minima in 
$P_{\mathrm{O}}$ appear at matching fields. $U^{\prime \prime }$ in a power sweep is shown for $f=1$ GHz in panel c). The right part are cross sections at fixed powers. All panels correspond to the Pb-PPC sample.}
\end{figure}

\section*{Discussion}

\textbf{Mechanism of nonlinear vortex response and modelling.} Mechanisms of nonlinear response of vortices to microwave
radiation are far from being fully understood. LO theory \cite{Larkin1975} predicts a nonlinear response at sufficiently large electric fields, that induces a high speed DC motion of the vortices. If their speed exceeds some critical value, $v_{c}$,
much lower than the critical velocity for breaking Cooper pairs,
the current decreases with increasing voltage. This is a consequence of an electronic instability of the non-equilibrium distribution of quasiparticles
at high velocities, leading to a reduction of the vortex core size.
A further increase of vortex velocity leads to an abrupt switching into a
state with higher electric resistivity. On the other hand, the nonequilibrium quasiparticle distribution
close to the energy gap where the density of states is maximal can cause
stimulation of superconductivity \cite{Eliashberg1970}. One can anticipate
an interplay between the above mechanisms in a mixed state of \textit{mw} driven type-II
superconductors. However, we are not aware of a theory quantitatively
interpreting our experimental results.

\begin{figure}
\includegraphics[scale=0.3]{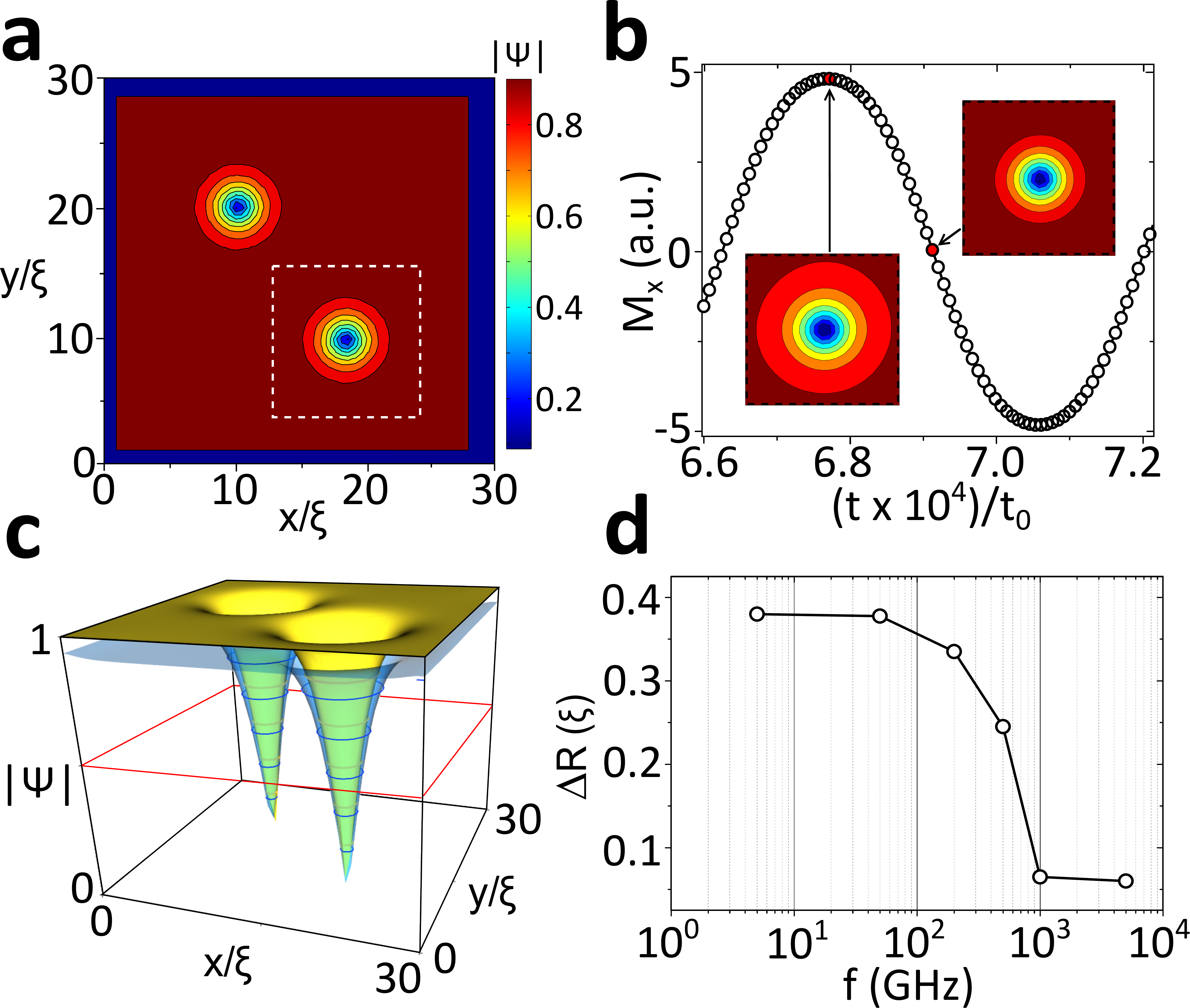}
\caption{Time dependent Ginzburg-Landau simulations.
a) Contour plot of $|\Psi|$ in a square sample, with $H_{DC}=0.02
H_{c2}$ perpendicular to the sample, after applying $H_{DC}=0.8 H_{c2}$ and
slowly reducing it. Two vortices are isolated. b) Oscillation of $M_x$ (component of magnetization in the direction of the ac field, in arbitrary units) as a
function of time, following the ac field at 5 GHz. Snapshots of the area
marked by dashed white lines in panel a) show the different size of vortices
at times separated by 1/4 of a period. c) $|\Psi|$ at the whole sample for
the two cases considered in b). Yellow surface corresponds to minimum and
blue to maximum of $|H_{DC}|$. d) Change of vortex radius (taken at $|\Psi|$%
=0.5, marked by red line in panel c) as a function of frequency in an
oscillation period. At high frequencies the vortices cannot follow the
external field and their shape and position almost don't change}
\end{figure}
To understand the nonlinear response of \textit{mw} driven vortices
we have simulated the \textit{ac} response of vortices using the time
dependent Ginzburg-Landau equations (see Suplementary material for details). The time derivative of the order parameter modulus $|\Psi |$ shows that vortices
oscillate about their equilibrium positions, especially at lower
frequencies, when they are able to follow the external \textit{ac} field
without delay (Supplementary Fig. A.9 b). As the \textit{ac} field
amplitude increases (which is equivalent to increasing the \textit{mw} power
of our measurements) the order parameter oscillates throughout the entire sample, being weaker at the maxima of amplitude of \textit{mw} field $h_{rf}$, as expected. This effect
is specially pronounced in the outer part of vortex cores, as can be seen in
Fig. 4b,c. When one compares the radius of a vortex (at a given
value of, for example, $\left\vert \Psi \right\vert =0.5$, see Fig. 4c) for different moments of an oscillation period, vortices are
narrower at zero \textit{ac} field amplitude than at its maximum \textit{ac}%
. As happens in vortex cores displacement, the higher the frequency of the 
\textit{ac} field, the more difficult is for a vortex core radius
to change size (see Fig. 4d). A transition from
linear to nonlinear response regimes is observed, \textit{leading to a substantial
reduction of the average vortex size} at high \textit{mw} drives as a
function of \textit{f }. The reason for the different ranges of frequencies considered in the experiment and simulation are commented in the methods section.

 The oscillations of the vortex core size under 
\textit{mw} radiation are always present, and more
notorious for higher \textit{ac} field amplitudes, which is in qualitative
agreement with the DC model by LO \cite{Larkin1975}. The above confirmation
of LO-type mechanism in \textit{ac} conditions agrees with simulations of DC driven vortices \cite{Vodolazov2007} and does not exclude
electron overheating in the vortex core as an additional factor contributing
anomalous velocity dependence of vortex viscosity \cite{Gurevich2008}.

The vortex velocity can be limited by the critical
value $v_{c}$ for the LO instability~\cite%
{Larkin1975,Doettinger1994,Silhanek2012}. Assuming that the maximum \textit{mw}%
-induced shift of a vortex is limited by the inter-dot distance $a\lesssim
(1-1.5)$~$\mu $m and that the dependence of $P_{\mathrm{O}}$ on $f$ starts to
saturate at $f\gtrsim f_{\mathrm{sat}}\sim 10$~GHz we estimate the
geometrically restricted maximal vortex velocity as $v_{\max }=f_{\mathrm{sat%
}}\cdot a\approx 6-10$~km/s. This is 2-3 times larger than the values of $v_{c}$
reported for Nb and high-$T_{c}$ superconducting films~\cite%
{Grimaldi2009,Wordenweber2012}.

\textbf{Summary and conclusions.} The experimental observation of stimulated superconductivity in type II superconductors has been used to quantify relative changes in vortex dissipation as a function of mobility (pinning). At high enough \textit{mw} power and/or low enough \textit{mw} frequencies (Fig. 3c)\ when MSSC is not
effective, the vortex matching effects are clearly observed as
periodic \textit{dips in the mw losses} when $H_{\perp }=n\Phi_{0}/a^{2}$ ($n$ is
an integer number and $\Phi_{0}$ the magnetic flux quantum). In contrast to
that, in a broad range of \textit{mw} powers (sufficiently below limiting
values which heat the sample) and at high enough frequencies (above about 0.6 GHz)
mobile (off-matching) vortices dissipate less than pinned vortices. One clearly
observes \textit{peaks in the vortex dissipation} in matching conditions. The higher the frequency, the broader the 
\textit{mw} power range where matching anomalies are seen as peaks in losses. Microwave stimulation changes from dips at matching fields at the lowest frequencies to
peaks at frequencies exceeding a few GHz, in agreement with TDGL simulations, that indicate a transition to a nonlinear regime when mobile (interstitial) vortices dissipate less than pinned ones. The observed effects (transition from peaks to dips) remain qualitatively unchanged for the range up to 3 vortices per pinning center, meaning that intervortex interaction has a weak influence on our resuts. The supplementary video 1 shows a simulation of the vortex response to a \textit{mw} magnetic field. One clearly observes the changes of the vortex core radius and (through differential analysis of the modulus of the order parameter) the vortices motion. We find vortex deformation to be minimum because its displacement at \textit{mw} frequencies is small in comparison with radius oscillations.

We point out that stray fields of Py dots do not play an essential role for the effects we observe. Unlike previous simulations \cite{Milosevic2004}, our dots are in the magnetic vortex state with minimum stray fields \cite{Gomez2013} and with ferromagnetic resonance (FMR) suppressed \cite{Aliev2009}, i.e., the dots are not saturated. Besides, there should be a strong structural pinning profile due to the fact that the SC film covers the array of dots and not vice versa as in \cite{Milosevic2004}. A qualitative similarity in the microwave losses measured with perpendicular or with inclined nearly parallel magnetic fields (Supplementary Fig. A.8) further confirms that the observed effects are not induced by the presence of an in-plain component of magnetic field.

In conclusion, we have observed experimental signatures of stimulated superconductivity in type II superconductors in the vortex state including an enhancement of the upper critical fields and a somewhat less noticable increase of the critical temperature. Moreover, we have found experimentally and supported by simulations the unique fingerprint of MSSC in vortex dynamics -the reduced dissipation of microwave driven vortices due to a reduction of the vortex core size.
Besides significance for condensed matter physics, our results may have implications for the current controversy on the possibility of stimulated superconductivity in holographic superconductors \cite{Enhanced_HS_1,Enhanced_HS_2}.

\bigskip

\section*{Methods}

We investigated two types of samples: plain 60~nm thick Pb films ($T_c \simeq 7.2 $K) and 60~nm thick Pb films deposited over a square array of periodic pinning centers (Pb-PPC), consisting of circular Py dots (see Suplementary material for further details).  All figures (except Fig. 2d and Supplementary Fig. A.5) refer to the Pb-PPC sample.

The broadband measurements were done with a Vector Network Analyzer (VNA) connected to a coplanar waveguide (CPW) situated inside a cryostat with a superconducting magnet (see Suplementary material for details). The VNA signal excites the sample, placed on the CPW (Fig. 1a). The complex \textit{mw} permeability, $U\equiv U^{\prime }+iU^{\prime \prime }$, is determined as
the VNA transmission parameter $S_{21}$, dependent on microwave power ($P$), frequency ($f$), temperature ($T$) and magnetic field ($H$), normalized by $%
S_{21}$ at a reference $H$ or $T$, in the normal state (Suplementary material for more details). Experimental figures corresponds to the estimated values of power waves travelling through the waveguide, but not absorbed by the vortex system.

To understand better the individual behavior of superconducting vortices under the influence of an in plane \textit{ac} magnetic field, we have simulated the TDGL
equation in 3D. Simulations allow to include a DC field perpendicular to the
film to create vortices, and a sinusoidal field parallel to the plane that
represents the \textit{mw} field generated by the CPW. Both field components
are introduced through the appropriate boundary conditions (see Suplementary material). Our simulations are based on the finite difference approach used
several times in the past in 2D (see for example \cite{Buscaglia}). The Ginzburg-Landau parameter used is $\kappa=2$. The temperature has been fixed far from $T_c=7.2 K$ ($T=4 K$) because vortices
are better observed, being the results obtained still valid (although less vissible) at higher temperatures.
A mismatch between the frequency range presented in the measurements and that of the simulations is due to the absence of precise knowledge of the characteristic time scales of the normal and superconducting parts of the GL simulation. We use our simulation just for qualitative confirmation of existence of LO mechanism for the microwave driven vortex.
Future work could also try to analyze numerically possible coupled magnetic dot-superconducting vortex dynamics. This task, however, presents great challenge because of the need to include dynamics of magnetic pinning centers.

\section*{Acknowledgements}
Authors gratefully acknowledge Yu. Galperin, T. M. Klapwijk and V. Vinokur for insightful discussions and A. Awad for experimental help on the initial stages. This work has been supported in parts by Spanish MINECO (MAT2012-32743), Comunidad de Madrid (NANOFRONTMAG-CM S2013/MIT-2850) and NANO-SC COST-Action MP-1201.  A. Lara thanks UAM for FPI-UAM fellowship. Authors thank CCC-UAM (SVORTEX) for computational capabilities. The work of A.V.S. was partially supported by “Mandat d’Impulsion Scientifique” of the F.R.S.-FNRS.

\newpage

\newpage
\appendix

\counterwithin{figure}{section}
\section{Supplementary Material}
\textbf{Experimental details:} The 60 nm thick Pb films were
electron-beam evaporated onto liquid nitrogen-cooled Si/SiO$_{2}$
substrates. All samples were covered with a 20 nm thick protective layer
of amorphous Ge, to avoid oxidation and keep them from getting scratched, since they are placed
face down over the CPW. The Pb film is evaporated in all samples over a
surface of 5$\times $5 mm$^{2}$. In the Pb-PPC samples, the Pb films were
deposited over a 2$\times $2~mm$^{2}$ square array of 30 nm thick 1000 nm
diameter Py dots, with an interdot distance of 2000 nm.

\begin{figure}[h]
\includegraphics[scale=0.5]{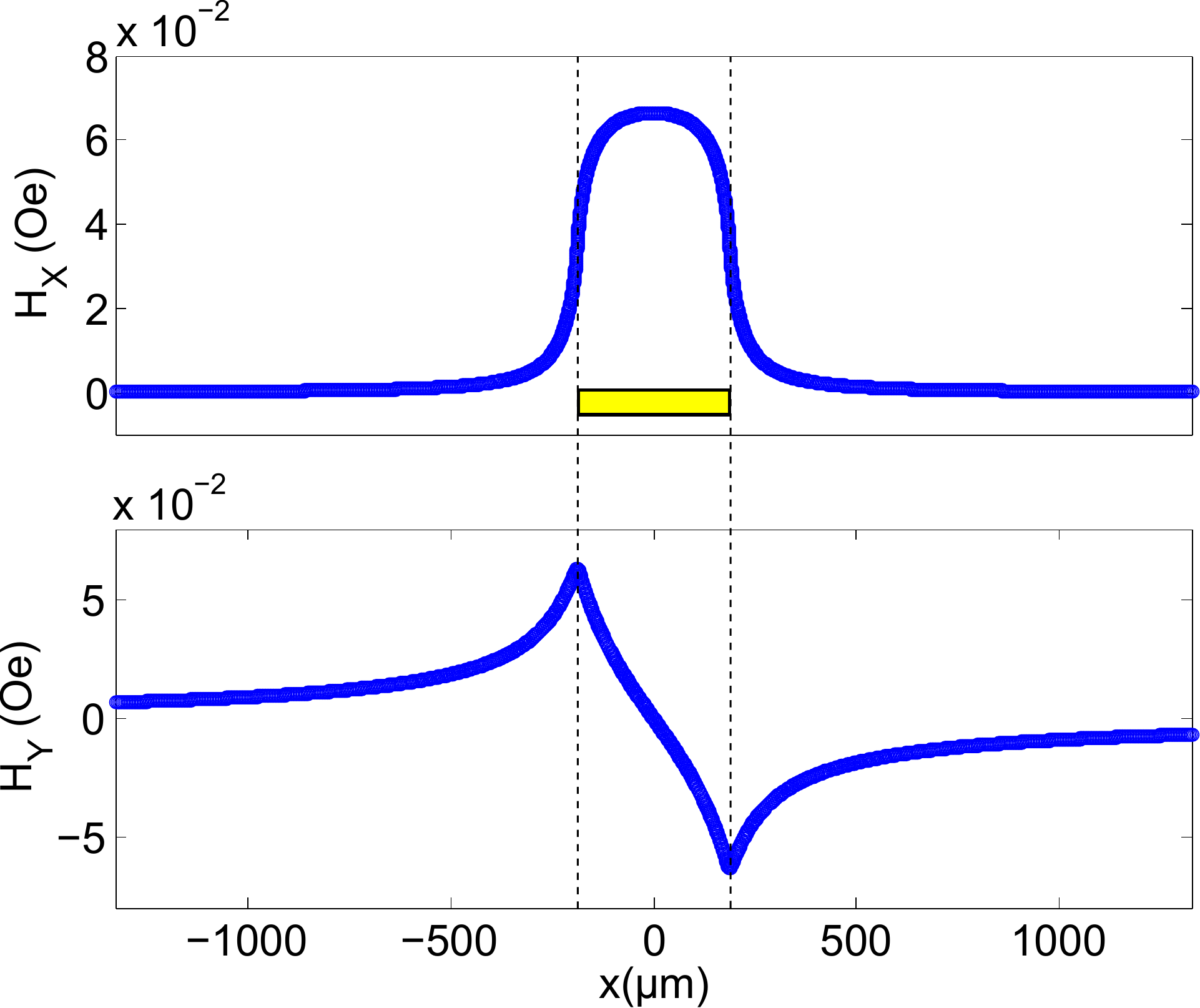}
\caption{Magnetic field generated by the CPW at a vertical distance of 20 nm (sample surface) and $P=5$ dBm. The yellow rectangle shows the dimensions of
the central conductor of the CPW.}
\label{field}
\end{figure}

 Hysteresis cycles confirmed that the Pb films are type II
SC, in accordance with previous reports~\cite{Dolan1973}. To reach the
superconducting state, a JANIS helium cryostat with a superconducting magnet
inside is used. Using two temperature control loops, the temperature
stability is better than 0.2 mK and could be maintained up to 3 days.%
\newline
The sample is placed at the end of an insert which contains a coplanar
waveguide (CPW) to provide the microwave drive field ($\mathbf{h}_{\mathrm{rf%
}}$) up to about 0.1 Oe parallel to the plane. Supplementary Fig. \ref{field} shows a calculation of the 
$x$ and $y$ components of the magnetic field generated near the central
conductor, at a vertical distance of 20 nm for $P=5$ dBm. The
expressions found in \cite{Chumakov} have been used for these calculations.

\begin{figure}[h]
\includegraphics[scale=0.5]{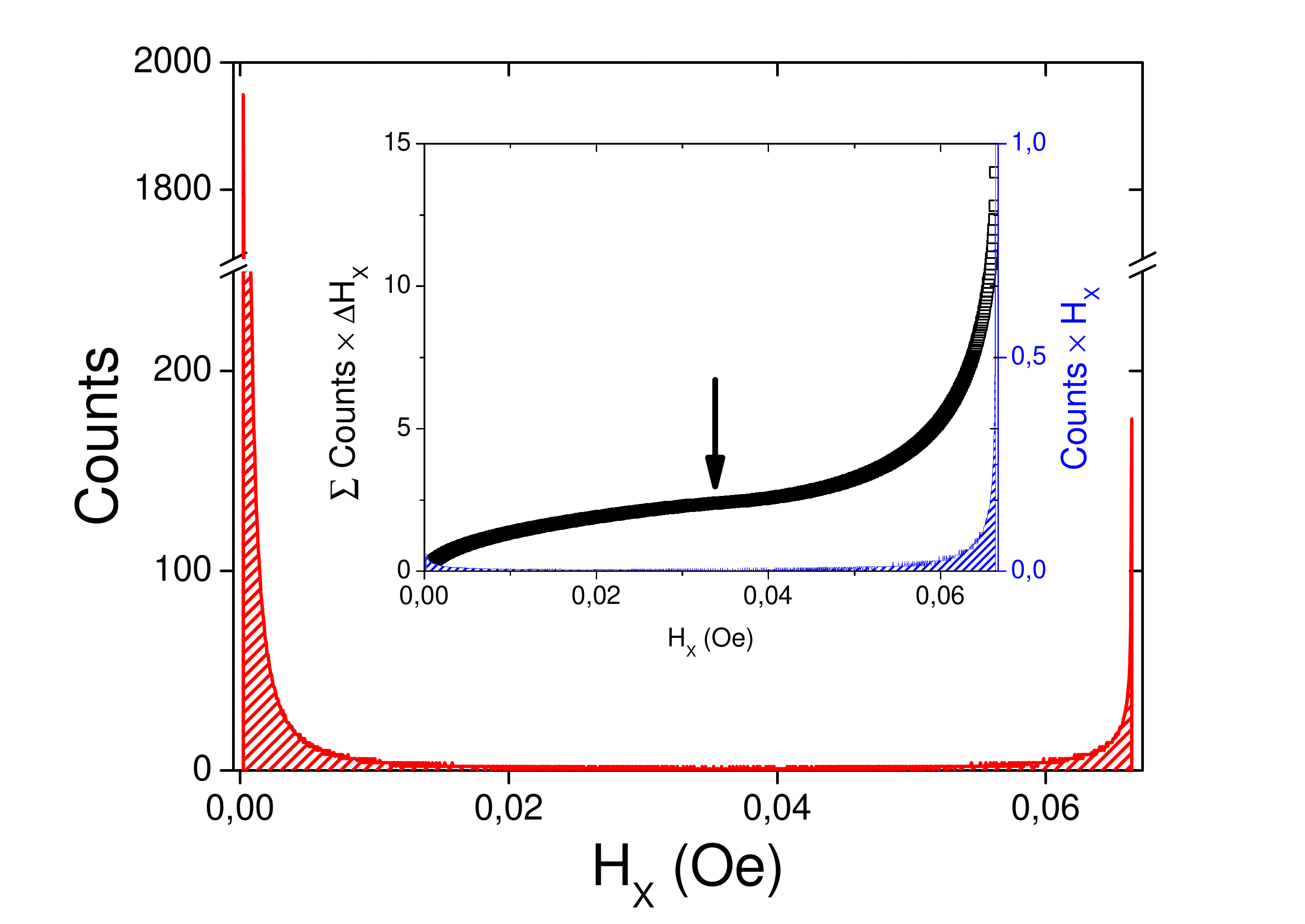}
\caption{Histogram of distribution of magnetic microwave field created by the CPW. The inset shows a weighted distribution of fields (blue area) and a cumulative distribution. The arrow marks the field at the edge of the central conductor.}
\label{histogram}
\end{figure}

The CPW is designed to have a characteristic impedance of 50 $\Omega $ up to
25 GHz and emits mainly in the TEM mode. The central conductor is 375 $\mu$m wide, and the gap to ground planes is 140 $\mu$m. The CPW is made of Rogers laminates RO4350 10MIL/RO4003C 8MIL with a copper gladding (66 $\mu$m thick), and an extra gold plating on top. This waveguide does not excite the sample (placed on top of it) with a homogeneous field. Therefore, the vortices in different parts of the sample will feel a different amplitude and direction of the excitation. To account for how much vortices are excited on top of the central conductor, compared to far from it, in Suppl. Fig. \ref{histogram} we show a histogram of the distribution of intensities of the microwave field. The largest value of field (corresponding to right in the middle of the central conductor) is found with a large frequency in the histogram, due to the almost parallel shape of the field close to this part of the waveguide. A large contribution can be seen in the histogram at low fields if long distances from the central conductor are considered in the calculations. With the appropriate weighting of each field with its frequency of appearance (inset of Fig. \ref{histogram}), we find that at a field $H_x$=0.034 Oe (corresponding to the edge of the CPW), this weighted signal equals $17\%$ of the maximum, which is achieved at the center of the CPW. Therefore, vortices lying on top of the central conductor will be excited by an $83\%$ of the total field delivered by the waveguide. The rest of the excitation lies beyond the central conductor.  

More details on this setup can be found in \cite{awad}. The high frequency
signal is emitted from one port of the Vector Network Analyzer, VNA, (an Agilent E8363C PNA model was used,
with frequency range up to 40 GHz and power up to 7 dBm) and propagates
through the CPW, to finally reach the other port. 

The VNA emits a signal of some known power, but due to reflections inside
the high frequency cables (two 1.5 m long cables in the insert to carry
the signal inside the cryostat, and two 0.6 m long cables to connect them
to the VNA) and connectors, and their heating due to eddy currents,
microwave power is not fully transmitted from one port of the VNA to the
other. The value of the signal frequency ($f$) is related to the appearance of
internal reflections, and a $f$ sweep gives the information of how much
power is transmitted at every frequency by means of the $S_{21}(f)$
parameter. The power of the signal at the CPW, that we refer to as $P$ in
the article, is estimated as the average of the power emitted from port 1
and the power received at port 2, since the cables configuration connected
to both ports is symmetric. We start from the definition of the
transmission parameter in terms of voltages, when no signal comes out from port 2: 
\begin{equation}
\left\vert S_{21}\right\vert =\left\vert \frac{ V_{2,in}}{V_{1,out}}\right\vert
\end{equation}%
In this expression, $V_{2,in}$ refers to the voltage (amplitude and phase)
of waves entering port 2, and $V_{1,out}$ to voltage of waves emitted from
port 1. In terms of power, the expression for the amplitude is: 
\begin{equation}
\left\vert S_{21}\right\vert =\sqrt{\frac{P_{2,in}}{P_{1,out}}}
\end{equation}%
Knowing $S_{21}(f)$, directly measured with the VNA, the power at port 2 is
calculated as: 
\begin{equation}
P_{2,in}=|S_{21}|^{2}P_{1,out}
\end{equation}%
and the estimated power at the waveguide, $P$, is taken as the average: 
\begin{equation}
P=\frac{P_{1,out}+P_{2,in}}{2}={P_{1,out}\frac{(1+|S_{21}|^{2})}{2}}
\end{equation}%
It is to be noted that this $P$ does not refer to the radiated
electromagnetic energy incident on the sample, just to the power associated to the current that
produces this radiation, as it propagates through the CPW.\newline
This correction helps to quantify the losses due to the relation between wavelength and length of the cables. Higher frequency signals suffer more reflections, and are not so well transmitted to the second port, which can be quantified via $S_{21}$. However, more factors can contribute to decrease the power that is delivered to the sample, such as not perfectly symmetric cables, that would decrease the accuracy of the previous average. Also, it is important to remember that working at low temperatures can result in important differences compared to room temperature, due to changes in the length of cables, etc. In our case, this is not a problem, since the range of variation of $T$ is small, always very close to $T_c$. Also, even if we could take into account all losses, the results presented in the main text are relative (throughout the range of powers considered, where transmission efficiency is invariant, since the frequency is fixed), and a correction of the absolute value of power will not affect the presence of the observed effects.\\

\begin{figure}[h]
\includegraphics[scale=0.3]{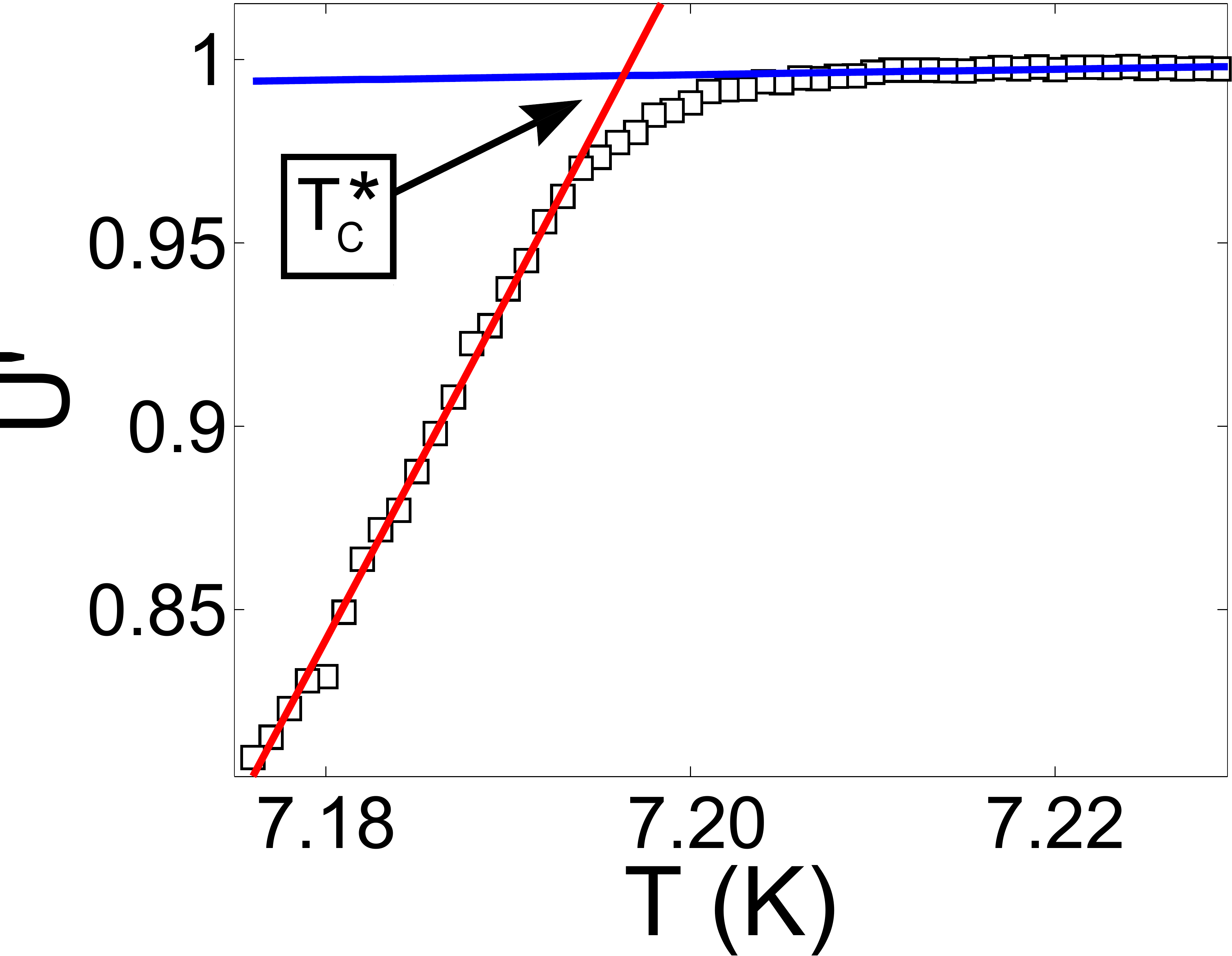}
\caption{Method to determine $T_c$. In this case, $U^{\prime }$ is
considered.}
\label{Tc}
\end{figure}

\textbf{Measurement of the permeability parameter U: } For the data analysis (i.e. calculation of the real
and imaginary parts of $U$) the reflected signal is neglected ($S_{11}$
coefficient of VNA), since its relative variation with respect to a reference trace is more than 15 times lower in magnitude than the
transmitted signal ($S_{21}$ coefficient). The same increase of $T_c^*$ and $H_{c2}^*$ at intermediate powers has been observed on the reflection coefficient $S_{11}$ alone as well.\newline
The permeability parameter was analyzed as the VNA transmission parameter ($%
S_{21}$) for every field normalized by the same parameter corresponding to
the normal state of the sample, similar to the analysis method used in \cite%
{kuanr}:

\begin{equation}
\mu \propto U(f,P,H)=\frac{S_{21}(f,P,H)}{S_{21}(f,P,H_{\mathrm{ref}})}\,.
\end{equation}%
Here $S_{21}(f,P,H)$ and $S_{21}(f,P,H_{\mathrm{ref}})$ are the  ($f$)
and ($P$) dependent forward transmission parameters at the applied field of
interest $H$, and the reference field $H_{\mathrm{ref}}$ (the maximum
applied field, higher than $H_{c2}$). This expression is used when $T$ is
kept constant and $H$ is changed. An analogous analysis is used in $T$
sweeps at constant $H$, but normalizing at a reference temperature. It is
necessary to achieve the normal state, either by increasing $T$ or $H$, to
have a clear signal corresponding to the SC response, when compared with the
normal response, so well above $T_c$ or $H_{c2}$ conditions are necessary
for a correct normalization.  The quantity $U$ is complex, and the imaginary
part of $U=U^{\prime }+iU^{\prime \prime }$, $U^{\prime \prime }$,
represents microwave losses, while the real part $U^{\prime }$ represents
energy stored and exchanged between sample and circuit (CPW), in this case flux screening by the superconductor.\newline
Once $U$ is found, values of $T_c$ and $H_{c2}$ can be found when reaching
high enough values of $T$ or $H$, so that the region in which the sample is
in the normal state is clearly differentiated from the superconducting
region. Since the transition is not perfectly sharp and well defined, the
following method to determine $T_c$ and $H_{c2}$ will be used. To our best knowledge, there is not a universal criterion for extracting from permeability data the critical values of field and temperature. Therefore, in the following we will refer to the critical values obtained as $T_c^*$ and $H_{c2}^*$, that may not be the exact values of critical $T$ and $H$, but from which a clear dependence with microwave power is observed, that we interpret as stimulation of superconductivity by microwaves.\\
 $T_c^*$ is
extracted from $T$ sweeps at constant $f$ and $H$, but varying $P$, so that
MSSC can also be detected in these measurements. On the other hand, $H_{c2}^*$
is found from $H$ sweeps in which $T$ and $f$ are kept constant, while $P$
also changes, to evidence the presence of MSSC not only from the values of $%
T_c^*$, but also from $H_{c2}^*$. 
For the results of $T_c^*$ and $H_{c2}^*$ discussed in this article,
the real part of $U$ was used, only because its changes are larger and easier to
observe and analyze, but the superconducting transition is equally visible
in $U^{\prime }$ and $U^{\prime \prime }$. \newline
\newline
\emph{Critical temperature: }The values of $T_c^*$ were found in $T$ sweeps
by finding the intersection between piecewise polynomial fit of the response in
temperatures for each power in both the superconducting part, and the normal
part (see Suppl. Fig. \ref{Tc} for an example of a $T$ sweep at $f= 6$ GHz
and $P=-17$ dBm, corresponding to Fig. 1c in the article). The resulting
dependence of $T_c^*$ on power clearly shows an increase at intermediate
values of power. Analogous results (increase of $T_c$ at intermediate
powers) are also achieved when directly plotting the temperature at which
the response changes in a (for example) $1\%$ with respect to the normal state. The error in $T_c^*$, of 1 mK, is obtained statistically as the dispersion in the intersection, found by least squares. 

\begin{figure}[tbp]
\includegraphics[scale=0.7]{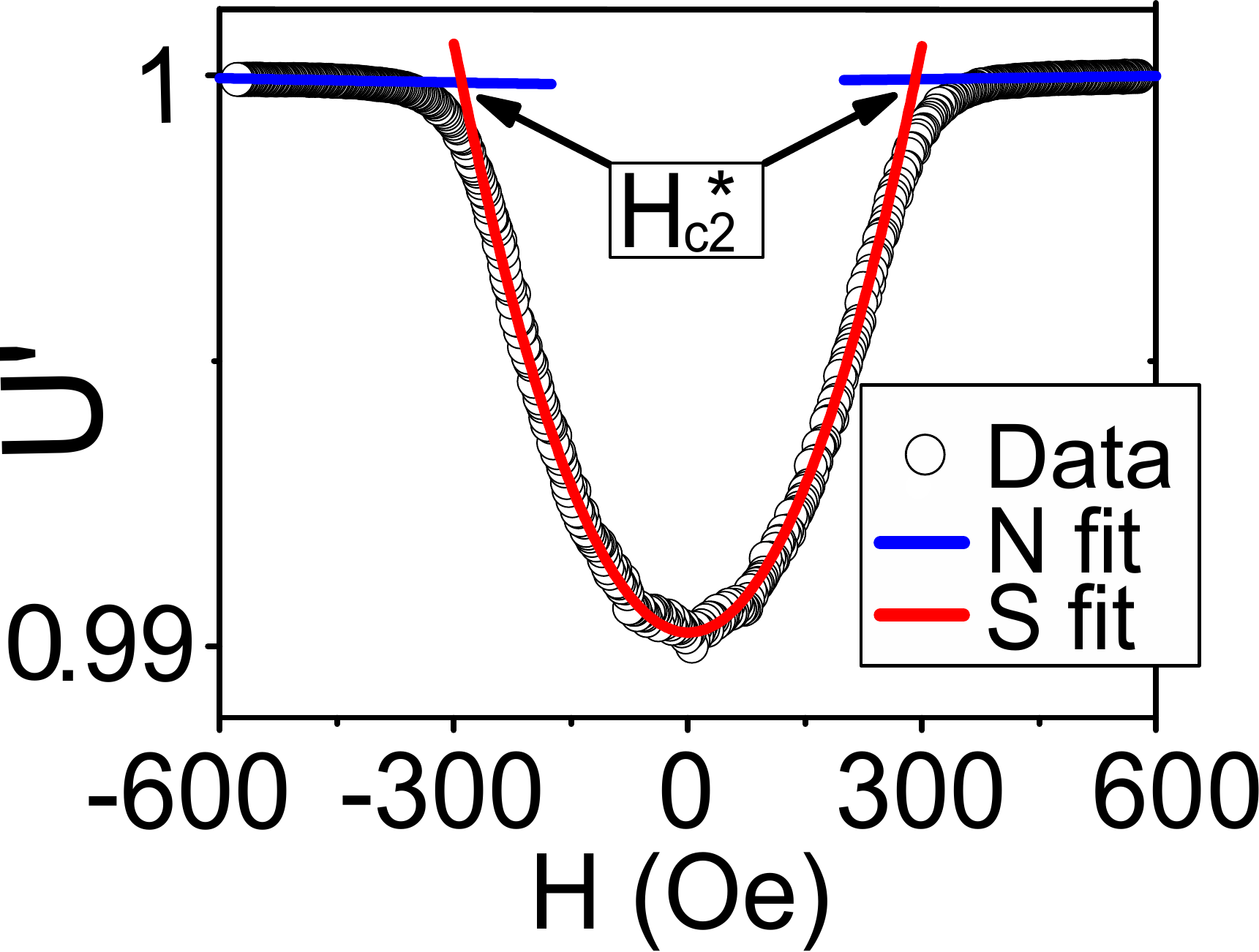}  
\caption{Method to determine $H_{c2}^*$}
\label{HC2}
\end{figure}

\emph{Critical field: }From measurements of $U(f,P,H)$ similar to those
shown in Fig. 2a of the article ($H$ sweeps) one can determine $H_{c2}^*$ as
the borderline separating the normal region from the superconducting one. As
can be seen in Suppl. Fig. ~\ref{HC2}, showing a typical magnetic field
dependence of $U^{\prime }(f,P,H)$ at $P=-15$ dBm, $T=7.17$ K and $f=6$ 
GHz while changing the applied field, the transition from normal to
superconducting state is not completely abrupt (the same happens when
finding $T_c^*$). We define $H_{c2}^*$ as the intersection of the polynomial low
field fit of $U^{\prime }(f,P,H)$ shown by a red solid line and the high
field level corresponding to the normal state and fitted by blue lines.

The value of critical field for each microwave power is considered as the
average of the values obtained for positive and negative fields, to cancel
the shift produced by the magnetic field frozen in the superconducting
magnet.\\
Analogous results (increase of $H_{c2}^*$ at intermediate
powers) are also achieved when directly plotting the field at which
the response changes in a (for example) $0.01\%$ with respect to the normal
state.\newline

\begin{figure}[h]
\includegraphics[scale=0.5]{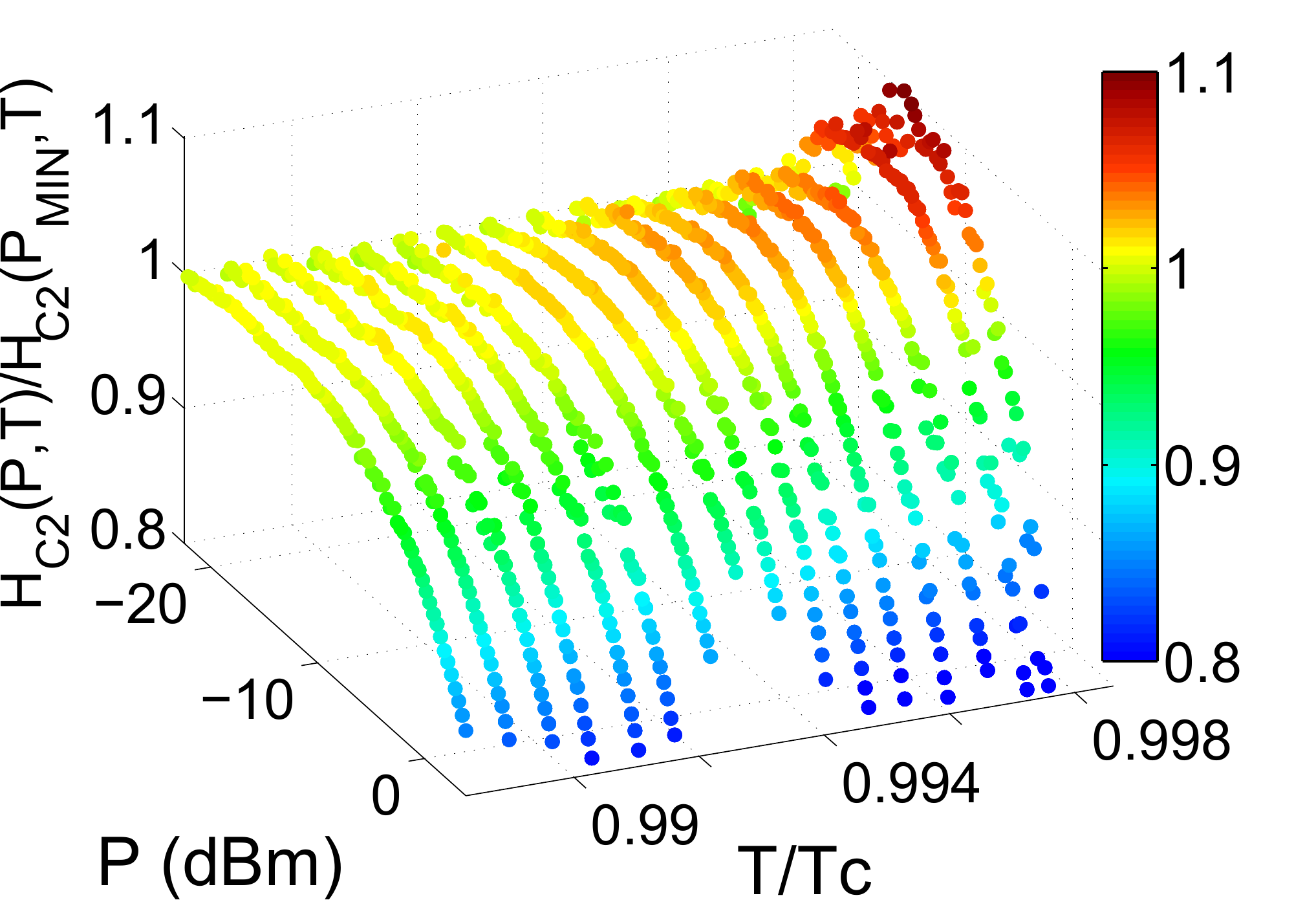}  
\caption{Relative critical field in the plain 60 nm thick Pb film}
\label{HC2rel}
\end{figure}

\textbf{Additional results and discussion: }\newline
\emph{Range of power considered:} The VNA that we used can sweep power down to -40 dBm. We show results from -27 to 7 dBm. At powers lower than -27 dBm the sensitivity starts to decay rapidly, and there is noise that does not allow to have clear results. Also, a change of range is at -27 dBm, and some discontinuities may appear in the response. Despite the noise, the tendency at lower powers (not shown) is to present a constant signal (no changes of $P_O$ are observed in that region).
\newline
\emph{Influence of pinning:} The article deals mainly with the sample with
PPCs, since the stimulation of superconductivity appears to be stronger than
in plain films. As a comparison, the same analysis presented in Fig. 2c in
the article is shown in Suppl. Fig. \ref{HC2rel} for the plain Pb film. It
can be noted that the maximum relative upper critical field is lower in the
plain films in comparison with Pb-PPC films. The introduction of pinning in
Pb-PPC therefore seems to enhance the MSSC-related increase of $H_{c2}^*$. The
experimentally observed stronger (and in a wider temperature interval)
effects in the films with artificial vortex pinning could be linked with
pairbreaking effects {\cite{Ducan1997}} from the Py dots stray fields \cite%
{Lange2003}

\begin{figure}[tbp]
\includegraphics[scale=0.15]{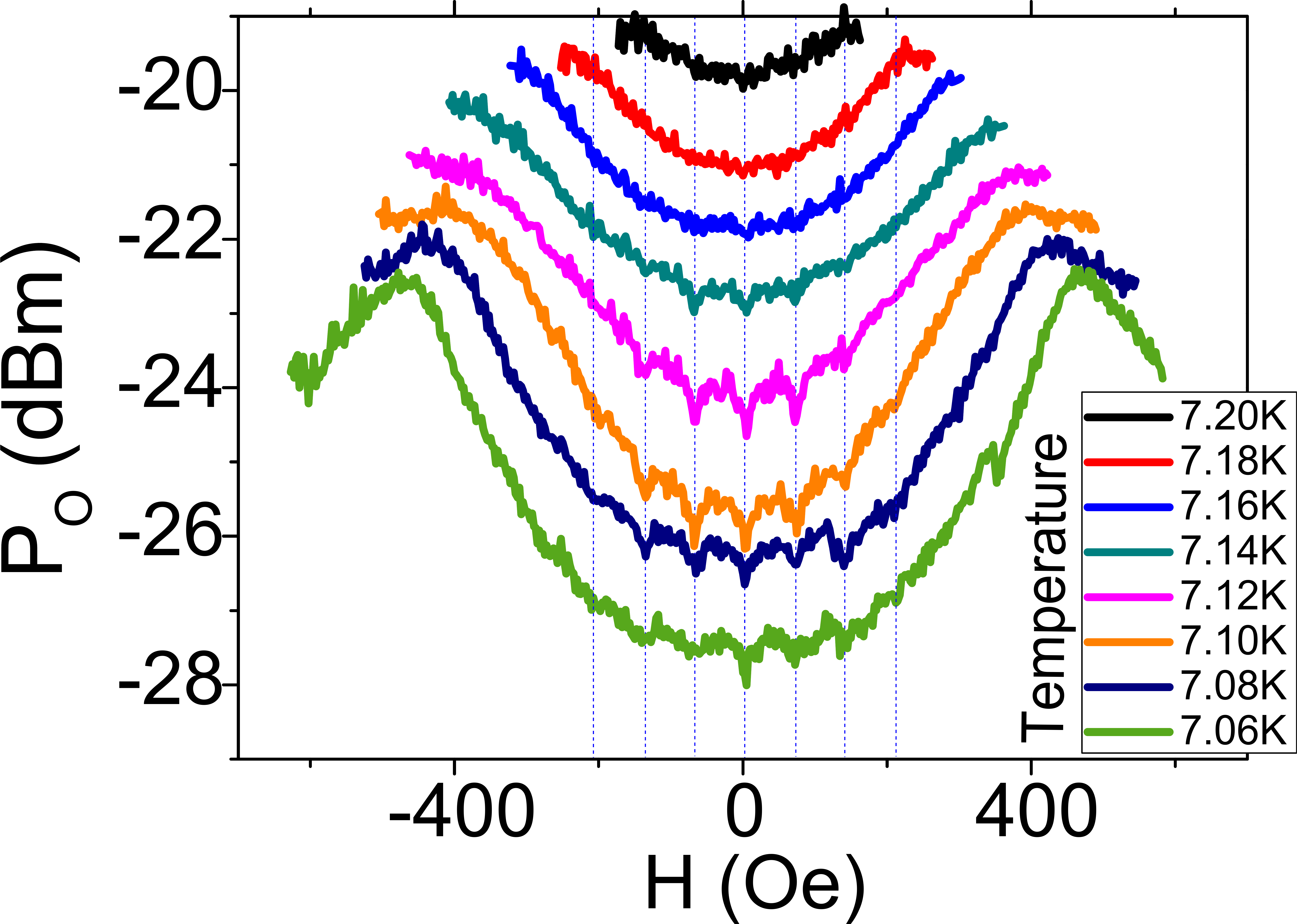}  
\caption{Optimum power as a function of magnetic field in the Pb-PCC film}
\label{Po}
\end{figure}

\emph{$P_{O}$ close to $H_{c2}^*$ far from $T_c$:} Concerning the dependence
of $P_{O}$ on $H$, as explained in the article, it increases when increasing 
$H$. Close to $T_{c}$ this trend is uniform, but at lower temperatures, and
higher fields local maxima in optimum power appear (see Suppl. Fig. \ref{Po}%
), as mentioned in the article, but not shown in Fig. 3b. We explain this
behavior as a consequence of a dense vortex structure, when vortices are not
able to move as freely as in cases with less density of them, diminishing
the LO effect. 

\emph{Anomalies in dissipation at matching fields:} At frequencies below 2 GHz it is observed that matching anomalies transform from maxima to minima
depending on $f$ and $P$, at $T=7.1$ K, as shown in Fig. 3 c) and Supplementary Fig. \ref{lowf} in the article.

In Figs. 3b and 3c in the main text and in Supplementary figures A.6 and A.7, matching fields have been indicated (\textit{a postriori}) with doted lines. From these values we can verificate the interdot separation of the dots comprising the array of PPCs, as follows:

$$
H_{\perp}=\frac{\phi_0}{a^2}\rightarrow a=\sqrt{\frac{\phi_0}{H_{\perp}}}
$$
Our measurements in the case of field perpendicular to the film (Supplementary Fig. \ref{par_perp}) give a value of matching fields of 4.8 Oe. With the standard value of the flux quantum, $\phi_0=2.06783\cdot 10^{-15}$ Wb. Then, the interdot distance is, as expected:

$$
a=\sqrt{\frac{\phi_0}{H_{\perp}}}=2.075 \mu m\simeq 2 \mu m
$$

Mainly the out of plane component creates the vortices that will give raise to matching conditions. Then, we can estimate the inclination of the field from the matching conditions. We know that matching fields for a perpendicular field occur at 4.8 Oe. Then, if we measure them at 72 Oe in the inclined configuration, as in Fig. 3b and 3c in the main text, we can extract the inclination angle as its sine, which equals the ratio of both quantities: $\sin \alpha = \frac{4.8}{72}=0.066 \rightarrow \alpha = 3.78 ^\circ $\\

\begin{figure}[h]
\includegraphics[width=0.8\columnwidth]{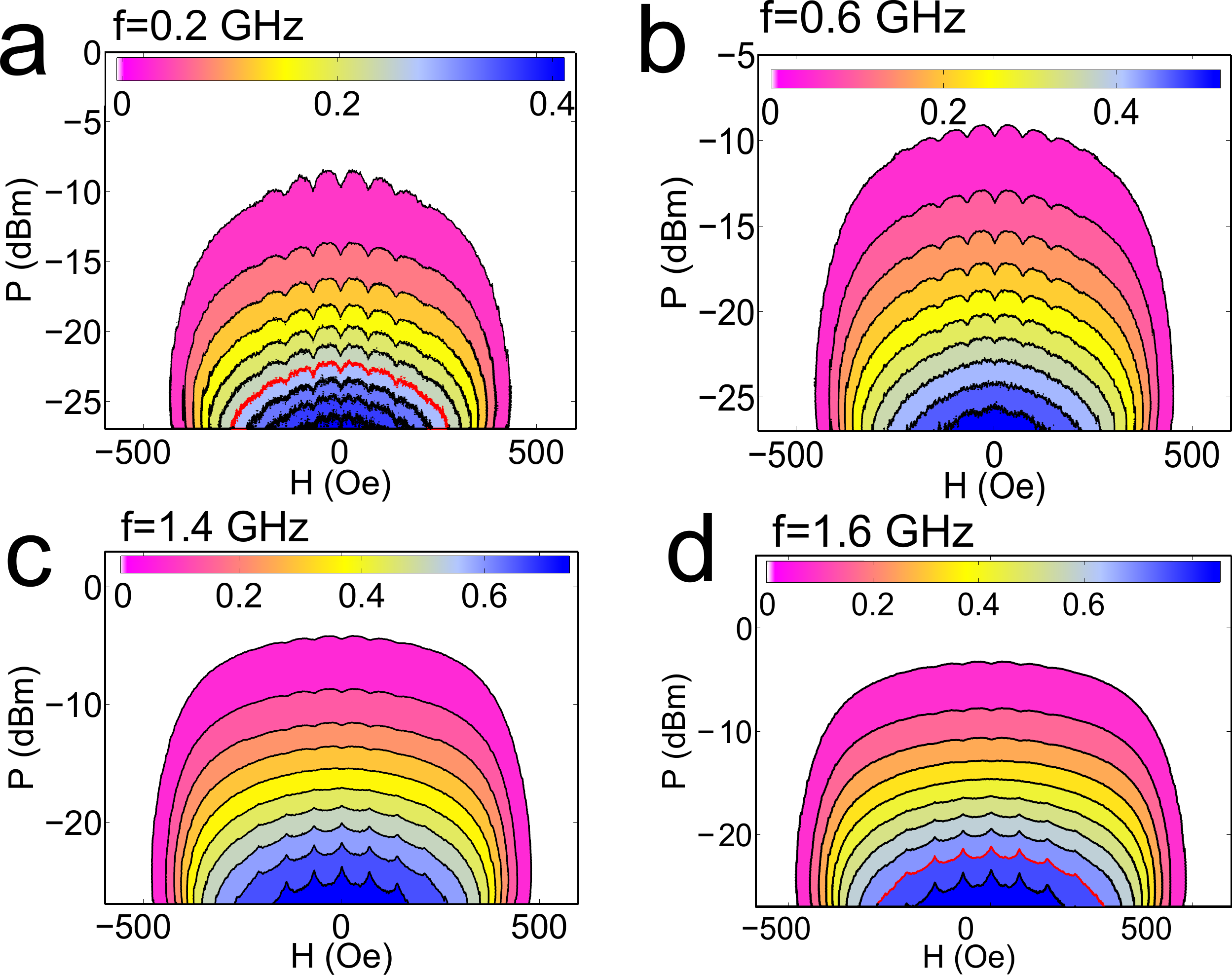}  
\caption{$U^{\prime \prime }$ as a function of $H$ and $P$ at different frequencies around 1 GHz. Panels a) and d) highlight a contour in red to stress the change from peaks to dips with frequency.}
\label{lowf}
\end{figure}

\emph{Dependence of the superconductivity stimulation on the magnetic field
direction:} Let us first discuss possible reasons why
MSSC in the type II superconductors could be more clearly observed in an
inclined magnetic field. We explain this observation as follows: As long as
the applied microwave frequencies are substantially less than corresponding
to the zero temperature gap values ($f_{Pb}(T=0 \mbox{K})=653$ \mbox{GHz}), and microwaves interact only
with normal quasiparticles, the effectiveness of MSSC is conditioned by the
presence of low energy normal quasiparticle states \cite{Eliashberg1970}. An
additional (in-plane)\ external magnetic field effectively suppresses the
order parameter, creating a gapless state in a rather extended field
interval $\frac{H_{c2}}{2}<H<H_{c2}$ \cite{Ducan1997}. Then, for the
suppressed order parameter values (i.e., close to $T_c$) the %
\removed{normal} quasiparticles DOS varies roughly as $\sqrt{E}$ (here $E$
is measured from the center of the gap) in the presence of impurities,
providing effective quasiparticle excitation by subgap (GHz)\ microwaves 
\cite{Ducan1997}. Our experimental observations (Suppl. Fig. ~\ref{Po})
reveal a clear enhancement of the MSSC strength in the form of an increase
of $P_{\text{O}}$ in the field interval ($\frac{H_{c2}}{3}<H<\frac{3}{5}%
H_{c2})$ followed by some small suppression of the $P_{\text{O}}$\ above $%
\frac{3}{5}H_{c2}$. Within our model, this increase is related in parts with
an increase of the low energy quasiparticles DOS in the indicated field
interval (mainly from in-plane field component) and enhanced contribution
of the LO effects (from perpendicular field component) while
vortex-vortex interaction remains unimportant.\newline

It could be argued that the in-plane component of the inclined field can produce changes in the stray field of the dots, that would in turn affect the behavior of vortices. Experimental data of the dots shows that their annihilation field is higher than the values we work with in this case, so the stray field is always kept to a minimum. Direct measurements of $U''$ comparing the case of perpendicular and inclined field confirm that matching fields fall in the same values (obviously, after correspondingly scaling the fields, to match the perpendicular to the plane components, Supplementary Fig. \ref{par_perp})

\begin{center}
\begin{figure}
\includegraphics[scale=0.1]{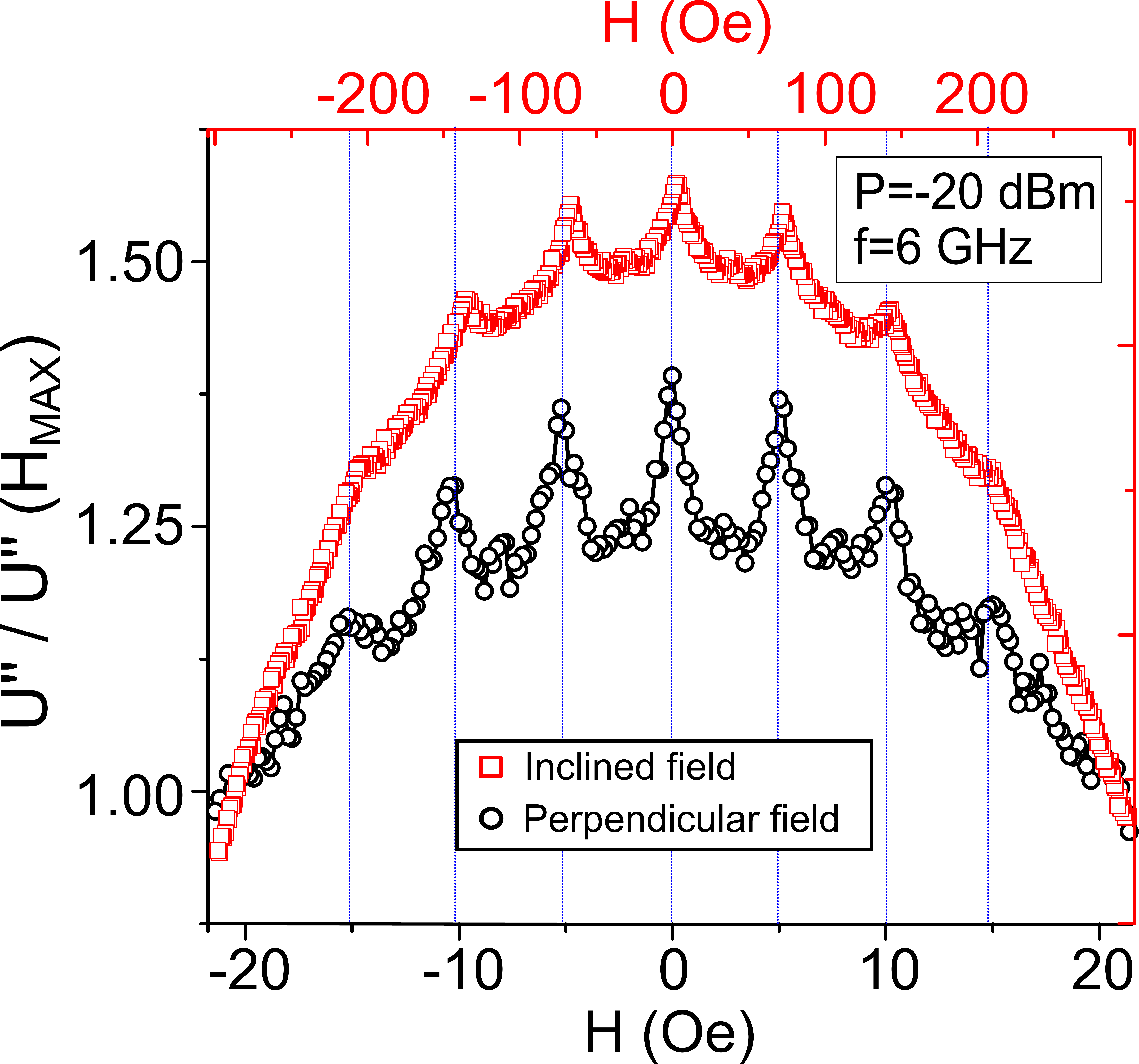}
\caption{Comparison of matching fields in the case of perpendicular and inclined field.}
\label{par_perp}
\end{figure}
\end{center}

\textbf{Time dependent Ginzburg Landau simulations: }\newline
In order to understand the behavior of vortices under an in plane \textit{ac} magnetic field, we have developed a computer program to simulate the 3D time dependent Ginzburg Landau (TDGL) equation. The equations can be expressed in a dimensionless form as follows \cite{Buscaglia}:

\begin{equation}
\frac{\partial \Psi}{\partial t}=-\frac{1}{\eta}\left[  \left( -i\vec{\nabla}-\vec{A}  \right)^2 \Psi +\left(1-T \right) \left( \left| \Psi \right|^2 -1 \right) \Psi \right]
\end{equation}

\begin{equation}
\frac{\partial \vec{A}}{\partial t}= \left( 1-T \right) \mbox{Re} \left\{ \Psi^*\left( -i\vec{\nabla}-\vec{A} \right) \Psi  \right\} -\kappa^2 \vec{\nabla} \times \vec{\nabla} \times \vec{A}
\end{equation}
 
Where $\Psi$ is the order parameter, $\vec{A}$ the vector potential and $\eta$ a constant relating the relaxation times of normal and superconducting electrons.
We use a finite difference scheme, a 2D version of which has been extensively used in the past. However, the more complicated geometry we want to reproduce cannot be achieved in 2D, since magnetic flux (used for boundary conditions) needs at least two layers to be accomodated in the simulated domain (flux enters through a surface, therefore a single layer in 2D is not enough for considering in plane component of the magnetic flux, just perpendicular to the plane).\\
The equations integraged in time are four, one for the order parameter (first Ginzburg Landau equation), and the other three for the auxiliar variables known as link variables (see \cite{Buscaglia} for more details), that can be obtained by rearranging the three components of the second Ginzburg Landau equation for the vector potential. Link variables are defined as:
$$ U^x_{x,y,z}=e^{-i\int^{x}_{x_0}A_x(\xi,y,z,t)d\xi}$$
$$ U^y_{x,y,z}=e^{-i\int^{y}_{y_0}A_y(x,\eta,z,t)d\eta}$$
$$ U^z_{x,y,z}=e^{-i\int^{z}_{z_0}A_z(x,y,\zeta,t)d\zeta}$$
Where $x_0$, $y_0$ and $z_0$ are arbitrary points of space that eventually cancel out. Given a plane, $XY$ for example, the rectangle formed by the cells in ($i,j$), ($i+1,j$), ($i,j+1$) and ($i+1,j+1$) will hold a circulation of the vector potential that can be related to magnetic flux at that position inside the superconductor:
$$ \oint _{\partial \Sigma}\vec{A} \cdot d \vec{l}=\iint_{\Sigma} \vec{B} \cdot d\vec{s}=\Phi _B$$
From such a rectangle the discretized distribution of magnetic field can be found in the $x=i$, $y=j$, $z=k$ coordinates:

$$U^x_{i,j,k}U^y_{i+1,j,k}\overline{U^x}_{i,j+1,k}\overline{U^y}_{i,j,k}=e^{-iB(i,j,k )\Delta x \Delta y}$$

and a similar calculation for the other two planes $YZ$ and $ZX$.

The geometry that we simulate is that of a thin film, with four cells in vertical direction. Doing so, magnetic flux is imposed according to boundary conditions (continuity of the parallel component of $H$) in the top and bottom planes, and allowed to evolve according to TDGL in the center of the film. The in plane dimensions are of $25 \xi \times 25 \xi \times 3 \xi$, being $\xi$ the coherence length, that is the unit of length of the simulations (its absolute value is not specified, it enters the computations through the Ginzburg Landau $\kappa$ parameter). This is enough to distinguish clearly vortices from each other and from the borders, and at the same time small enough to allow for relatively fast computations. This is so because 3D version of TDGL needs to consider terms that 2D does not, since there are more types of possible border cells, making it inevitably slower to calculate than just N times slower, being N the number of cells in vertical direction. 
Details about the discretization of TDGL equations for a rectangular mesh can be found in \cite{Buscaglia}.\\
By choosing an appropriate value of $\kappa=2$, vortices appear when applying a perpendicular to the plane magnetic field above $H_{c1}$. This field is applied until the vortex lattice is stationary (vortices enter from the borders towards the center). Then, the \textit{ac} magnetic field is applied parallel to the plane (keeping unchanged the perpendicular DC field). \\
Keeping track of macroscopic quantities (such as the total magnetization) we can extract information about the global response of the system (vortices and the rest of the system, including borders), such as difference of phase with respect to external signal (for very high frequencies, vortices start to not being able to follow the field). However, we are more interested in the ``microscopic" behavior of vortices, that can be studied by looking directly in the values of $|\Psi|$ at each simulation cell.\\
It has been found that vortices shape and position as a function of time depend both on frequency and amplitude of the external \textit{ac} field. In general, under an in plane \textit{ac} field vortices have been found to periodically shrink and widen in size (See Figure 4 in the main text). If microwave power (amplitude of the ac field) is large enough, these oscillations are seen very clearly. In Fig. 4d of the main text we show the variation in vortex core radius $\Delta r$ at a fixed value of $|\Psi|=0.5$ as a function of frequency for different powers. Higher powers allow to see better this difference, that is always present. The higher the difference of phase between \textit{ac} field, and dynamic magnetization is, the more difficult is for the vortex to follow the external excitation, and these oscillations are not so well observed.
As for the average redius of the vortex core, at higher frequencies it decreases, as predicted by LO, due to a redistribution of quasiparticles outside the vortex core.\\
\begin{figure}[h]
\includegraphics[scale=0.5]{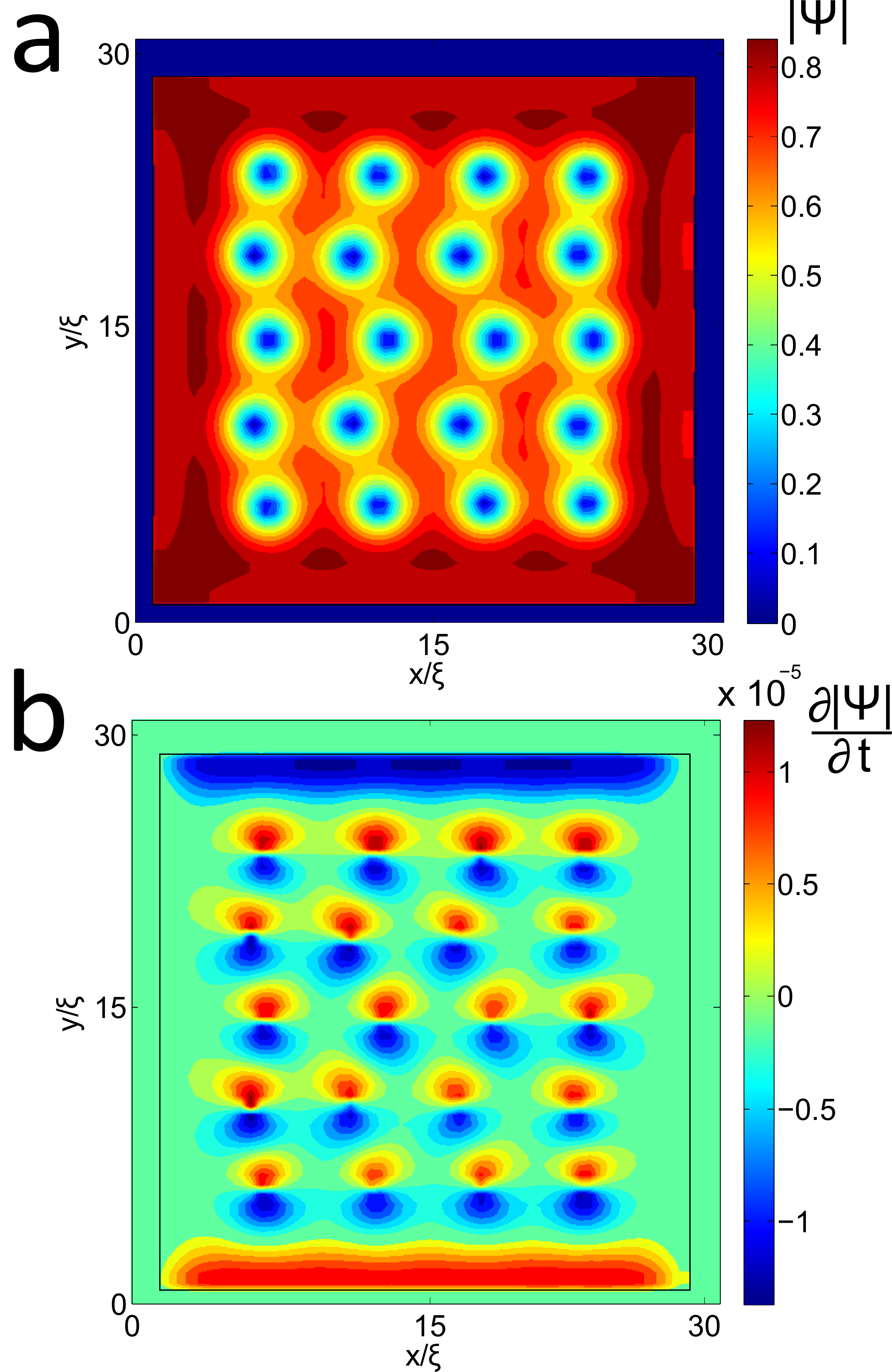}
\caption{a) Magnitude of the order parameter for $H_{DC}=0.4 H_{c2}$. b) Time derivative of panel a). Vortices move toward the blue area. Half a period later, colors interchange}
\label{simulations}
\end{figure}

It is to be noted that in solving TDGL equations, no energetic consdierations for stability of Cooper pairs is made, therefore effects such as Cooper pairs breaking due to absorption of photons of energy above the superconducting gap are not found. An arbitrarily high frequency \textit{ac} field just makes it harder for vortices to follow it, not being frequency a cause of loss of superconductivity. On the other hand, a higher amplitude of the \textit{ac} field does indeed destroy superconductivity. For example, at $|\mathbf{h}_{\mathrm{rf}}|=H_{c1}$, $|\Psi|<1$ far from vortices. Thus, not only vortices are affected by large amplitude ac fields, also the rest of the film (as was already discussed before, concerning the excess of normal quasiparticles (unpaired electrons) when applying an inclined DC field).\\
It is important to note that in TDGL equations, there is no notion of Cooper pair, neither of their bonding energy. Typically, photons of energy in the THz range are energetic enough to break a Cooper pair when being absorbed by them, since this energy corresponds to the superconducting gap energy $\Delta \sim $meV. Therefore, even when we apply an \textit{ac} field of frequencies the order of THz, the order parameter is not going  to drop to zero because we are reaching the frequency corresponding to the superconducting gap. This concept is not considered in the equations we are solving. TDGL equations are useful for our purpose in that they give information on the redistribution of normal quasiparticles by applying an \textit{ac} field to the vortex lattice, just by the deformation and movement of vortex cores. This redistribution of normal quasiparticles further interacts with the electromagnetic radiation of the waveguide, and create extra losses due to Joule heating.

\textbf{Influence of defects}
In the simulations, no periodic pinning centers are considered, we use them to study the reaction of vortices to a high frequency magnetic field.

However, the presence of defects can be considered. They are introduced by replacing the term $(\left| \Psi \right|^2 -1)$ by $(\left| \Psi \right|^2 -r)$ (see \cite{Buscaglia}). Here, the term \emph{r} is a variable that depends on position, and ranges between 1 and 0. By choosing values of \emph{r} lower than 1, the superconductivity is weaker at those points, and vortices are in a more stable position, so it acts as a pinning center for vortices. In simulations we have used a defect  of size $\xi$, with r=0 to trap a vortex. Several simulations at different frequencies show that regardless a vortex is inside a defect or not, the core size oscillates, as described in the main text. However, displacements of the vortex core are different, being more able to follow the external field when trapped. The vortices outside the defects also move but, especially at low frequencies, they tend to move in more circular trajectories, unlike vortices trapped in defects, which move more linearly. Also, radial oscillations of the vortex core size are observed both for vortices trapped inside defects, or outside defects, with a somewhat larger difference in amplitude at lower frequencies, as shown in Supplementary Fig. \ref{defect}. 

\begin{figure}[H]
\includegraphics[scale=0.5]{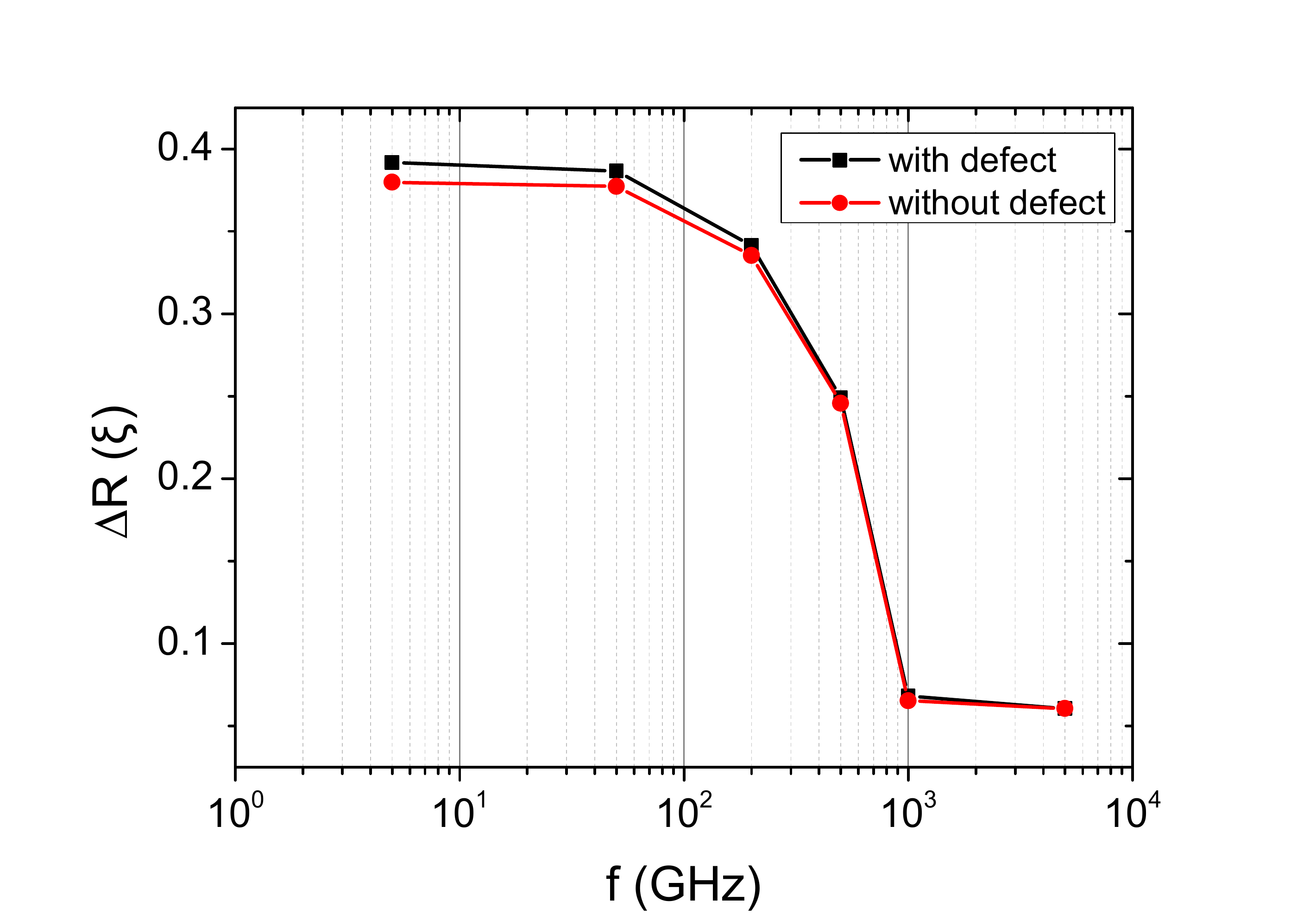}
\caption{Frequency dependence of radial oscillations magnitude for a vortex trapped in a defect, or outside of it.}
\label{defect}
\end{figure}

\end{document}